# Systematic Design and Demonstration of Multipole, Coupled-Cavity Integrated Photonic Bandpass Filters with High FSR, High-Q Single-Mode Microresonators in Low-Loss Silicon Nitride Platform

AMIR H HOSSEINNIA[1], ABDEL KARIM EL AMILI[2], YU-HUNG LAI[2], DANNY ELIYAHU[2], MOHAMMAD ENJAVI[1], LUTE MALEKI[2], ALI ADIBI[1,*]

[1]School of Electrical and Computer Engineering, Georgia Institute of Technology, Atlanta, GA, 30332 USA
[2] OEwaves Inc., 465 North Halstead Street, Suite 140, Pasadena, CA 91107, USA
*ali.adibi@ece.gatech.edu

**Abstract:** Tunable, low loss, narrowband, and frequency-stabilized filters have a critical role in realization of transceivers for efficient signal processing in telecommunications, and sensing applications at the presence of strong interference and noise. We systematically design, fabricate and demonstrate multi-cavity, multipole, integrated photonic bandpass filters with gigahertz (GHz) to sub-GHz bandwidth. The filters feature ultra-wideband (> 400 GHz) tunable center frequency, ultra-low insertion loss, steep roll-off, and compact footprint for radiofrequency (RF), microwave, and millimeter-wave (MMW) front-end applications. Using a set of design features for multi-cavity, high-order filters and supporting nanofabrication techniques in a silicon nitride (SiN) platform, the filters achieve record-high figures of merit and improve the state-of-the-art in integrated photonic filters for RF, microwave, and MMW systems. The resonator structures, the key feature for achieving the filter's performance parameters, serves as the building block with a combination of high FSR (~70 GHz) and high Q ($2.1 \times 10^7$). This is achieved through combining wide multi-mode waveguide segments and narrow single-mode regions that are connected with tapering. To the best of our knowledge, the experimentally achieved results of 520 MHz bandwidth, one FSR tuning range, 2 dB filter insertion loss, and up to 55 dB out-of-band rejection ratio are record high performance metrics for photonic filtering of RF, MMW, and terrahertz (THz) signals.

## 1. Introduction

Integrated photonic devices are highly attractive for generating, transmitting, and processing of ultra-high-frequency electrical and optical signals [1,2]. They offer the advantage of small size, weight and power together with compatibility for production at scale. Filters are essential in receiving, transmitting and processing of signals and play a major role in radio frequency (RF), millimeter wave (MMW), and terahertz (THz) systems, as critical components for separating signals from noise and improving signal to noise ratio (SNR) [3–5]. The advent of RF Photonic systems that support wide bandwidth and lower transmission loss underscores the need for In RF photonic systems, where higher the availability of bandwidth is a major advantage, photonic filters are essential in achieving improved noise performance and play an outsized role for direct up- and down-conversion of high frequency signals. While communication devices perform well at lower RF bands, applications such as satellite constellations demand high-rejection, tunable filters to mitigate crosstalk and detect weak narrowband signals distorted by noise or the atmosphere. This need is amplified in crowded frequency bands at microwave and mmW frequencies. The emerging 5G/6G systems for terrestrial and low earth orbit (LEO) networks [6,7], and advanced communication architectures such as orthogonal frequency division multiplexing (OFDM), pump isolation, and quantum communications, require high performance capabilities, beyond what is readily achievable.

Most frequency-selective photonic filters use linear effects, with some employing nonlinear phenomena like Brillouin scattering [8,9], Kerr effect [10-12], or optical frequency combs [13,14]. Linear designs typically exploit dispersive structures to convert wavelength-dependent phase shifts into amplitude filtering via interferometry, and require sharp low-loss dispersive elements. Common implementations include Mach-Zehnder interferometers (MZIs) [15,16], delay-lines [17,18], Fabry-Perot cavities [19], fiber Bragg gratings (FBGs) [20], optical microresonators (WGM or ring) [21–26], and arrayed waveguide gratings (AWGs) [27]. Microresonators are particularly suitable for realization of compact, tunable, high-selectivity integrated photonic filters. The ability to engineer their coupling and spectral response makes them ideal for scalable, cost-effective applications. While unit-cell-based, high-order digital architectures enable complex filtering, they face challenges in achieving steep roll-off, narrow bandwidth, low loss, wide tunability and robust performance. Achieving wide bandwidth, flat bandpass response, small footprint, sharp roll off, and high out-of-band rejection ratio simultaneously is challenging in conventional designs.

As a superior alternative, coupled-cavity systems—e.g., cascaded add-drop resonators—offer sharper roll-off, higher out-of-band rejection, and maximally flat (Butterworth) bandpass characteristics [22, 23, 28–30]. Despite impressive progress, the main challenges in using coupled-cavity filters are maintaining the needed precise control over inter-resonator coupling, suppressing fabrication-induced disorder, and mitigating excess insertion loss introduced by cascading multiple cavities.

In this paper, we design and demonstration of a novel five-pole, coupled-resonator, integrated photonic filter fabricated on silicon nitride (SiN) platform, with the ring resonators exhibiting combination of a high Q (~$2.1\times10^7$) with 70 GHz free-spectral range (FSR). The filter features efficient thermal tuning, and at telecommunication wavelengths (~ 1550 nm), has insertion loss as low as 2 dB, and out-of-band rejection beyond −55 dB. These performance parameters are achieved by careful design and fabrication of the ring resonator poles to ensure precise inter-cavity coupling while supporting various filter responses (e.g., Chebyshev, elliptic, Gaussian, Bessel) through parameter selection. As such, our results set a new benchmark in integrated photonic filter performance.

## 2. Design of Multipole, Narrow Bandpass Photonic Filters With a Coupled-Cavity Architecture

Figure 1(a) illustrates an example of a five-pole symmetric coupled-cavity optical bandpass filter used in this work. The main design parameters include the resonator shape and dimensions and the coupling coefficients between the input/through waveguides and the adjacent resonators

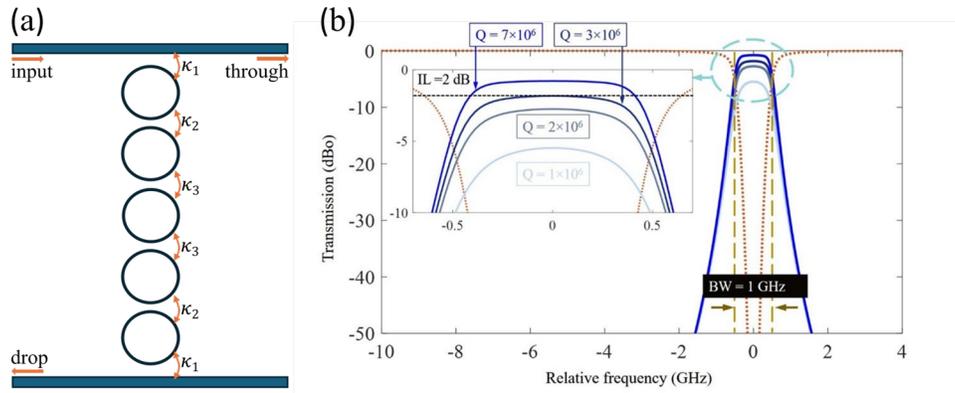

Fig. 1 (a) The schematic of a 5-pole microresonator-based integrated filter formed by identical cavities. The architecture has mirror symmetry across the middle cavity resulting in 3 independent coupling coefficients. (b) Theoretical optical power transmission spectra of a 5-th order Butterworth filter (light to dark blue: drop port; dashed red: through port) optimized for minimum in-band insertion loss (IL) while fixing the 3-dB passband at 1GHz. The curves show that a minimum Q of $3 \times 10^6$ is needed for IL = 2 dB. This values of Q for a mircroring resonator corresponds to an overall ~ 0.1 dB/cm optical loss for the corresponding waveguide. All simulations were performed using coupled-mode theory and implemented using MATLAB.

($\kappa_1$) and that between neighboring resonators ($\kappa_2$ and $\kappa_3$). Among different options, we use traveling-wave resonators in the form of modified single-mode microrings. As a parameter of design, FSR = 70 GHz is selected for the resonator to allow spurious-mode free and wideband performance, which is further discussed in the next section.

The coupling coefficients are found based on the desired filter response. We will focus on the Butterworth design for maximally flatband response [31,32]. The general formulas for designing a coupled-cavity filter for an N-th order Butterworth filter are [27]:

$$\kappa_i = \left(\frac{2\Delta_i}{1+\Delta_i^2}\right)^2, \Delta_i = \Lambda_i . \Lambda_{i-1}, \Lambda_i = \begin{cases} 1 & i = 0, N+1 \\ \sqrt{\frac{\pi f_{BW}}{2\theta_i f_{FSR_i}}} & i = 1,2,\ldots,N \end{cases}, \theta_i = 2\sin[(2i-1)\pi/2N], \quad (1)$$

where $f_{BW}$ is the 3-dB filter bandwidth, and $f_{FSR_i}$ represents the FSR of the $i$-th resonator. For example, in the structure in Fig. 1(a), $f_{BW}$ = 1 GHz results in $\kappa_1$ = 0.1357, $\kappa_2$ = 2×10$^{-3}$, and $\kappa_3$ = 6.2802×10$^{-4}$. Power efficiency, insertion loss, bandwidth, and out-of-band extinction ratio depend on waveguide losses and cavity quality factors (Qs). Figure 1(b) shows the simulatiuon results for the effect of the intrinsic Q of the resonator on the 5$^{th}$-order Butterworth filter response for the filter in Fig. 1(a), identifying Q ~ 3×10$^6$ (≈ 0.1 dB/cm loss) as the required value for achieving a target 2 dB insertion loss at 1 GHz filter bandwidth in this paper. Higher Qs also result in faster drop off of the filter response and higher out-of-band rejection ratios as discussed in more detailed in the Supplementary Materials (see Fig. S1).

## 3. Material Platform for Optimized Microresonator Performance and Spectral Response

To achieve the necessary high Qs, low-pressure chemical-vapor deposition (LPCVD)-grown SiN is used as an ultra-low-loss material [34–36] with 1–2 orders of magnitude lower than silicon (Si) in the near-infrared wavelengths. While highest Qs in WGM cavities for filtering can be achieved with large-scale microdisks and micro-donuts [33,37,36], their highly multi-mode nature significantly deteriorates the performance of coupled-cavity structures due to added coupling losses and spectral distortion. Thus, we use single-mode microring resonators to avoid higher-order modes and enable spurious-free responses and precise cavity-to-cavity coupling [37–38]. To reduce scattering loss, we reduce the waveguide thickness to 170 nm [39–41]. For single-mode cavities, the FSR must exceed the RF filter range (e.g., > 50 GHz for 0–25 GHz RF) due to double-sideband up-conversion. Yet achieving both high Q (> 5 × 10$^6$) and large FSR (> 10 GHz) in single-mode SiN cavities is challenging due to the need for larger radii of the microring to achieve higher Qs. While designing resonators using the Vernier effect can increase FSR, it degrades rejection ratio, complicates tuning, and increases power consumption [42,43]. Alternative high-aspect-ratio geometries, which can exceed 5 μm in width and have aspect ratios as high as 100 are lossy at bends, limiting FSR to 5–50 GHz while their large footprints limit dense photonic integration. In this work, we demonstrate, for the first time, single-mode microresonators with ~ 70 GHz FSR and record-high Q ~ 2.1 × 10$^7$ (at this level of FSR) enabling spurious-free, multi-cavity integrated photonic filters. This is achieved by engineering compact LPCVD SiN microrings that simultaneously maintain ultra-low propagation loss and small radii. Although we design the resonators for FSR = 70 GHz, the selected architecture allows FSRs potentially up to 200 GHz, nearly an order of magnitude higher than previous single-mode low-loss designs, while still preserving record-high Q.

## 4. Nanofabrication and Characterization of Integrated Photonic Microresonators Suitable for Coupled-Cavity Filters

The nanofabrication process for the coupled-cavity filter was developed in-house at Georgia Tech's Institute of Materials and Systems (IMS) with a focus on CMOS-compatible processes that enable mass manufacturing. It starts with a crystalline, undoped Si wafer, on which a thick thermal silicon oxide (SiO$_2$) layer is grown to isolate the optical mode and improve thermal performance. A high-quality stoichiometric SiN layer is then deposited using LPCVD, optimized for ultra-low optical loss. The wafer is coated with an electron-beam resist, hydrogen silsesquioxane (HSQ), followed by a thin conductive polymer layer (Espacer) to prevent charging during lithography. The device pattern is defined using an optimized electron-beam lithography (EBL) process and developed in a tetramethylammonium hydroxide (TMAH) solution. Pattern transfer into the SiN layer is achieved using an optimized fluorine-based inductively coupled plasma reactive ion etching (ICP-RIE) process. Post-etching, the wafer is annealed at a high temperature (1100 ºc) in a nitrogen environment to minimize surface and material defects. Finally, a top SiO$_2$ layer is deposited to symmetrize the optical mode and protect the SiN core. Characterization of waveguide-coupled single-mode microring resonators is done using a swept-wavelength transmission characterization setup near 1550 nm wavelength. A fiber polarization controller is used to adjust the input polarization for the characterization. Input/output focusing grating couplers are used to couple coherent light in and out of the access waveguides. Figure 2 shows the characterization results for a single-pole integrated photonic filter. The output light is detected using a variable gain photo-receiver and sent to a computer through a data acquisition (DAQ) card. Figure 2(a) shows raw data from a single-mode microring cavity (radius, R = 400 μm, 170 nm-thick SiN, 2 μm-wide waveguide) with a 2.5 mm roundtrip length. The waveguide–cavity gap, optimized via three-dimensional finite-difference time-domain (3D FDTD) simulations (Lumerical) for critical coupling, achieves a spurious-mode-free spectrum within each FSR, confirming the single-mode design. The measured FSR is 69.7 GHz, matching the 70 GHz designed target. Figure 2(b) magnifies a resonance in Fig. 2(a) near 1550 nm, showing an intrinsic $Q > 1.2 \times 10^7$—reportedly the highest achieved in a single-mode microring at this FSR in a CMOS-compatible SiN platform. Similar resonators with R = 350 μm and 300 μm (FSRs ~ 80 GHz and ~ 93 GHz, respectively) maintain $Q \geq 1.0 \times 10^7$. Note that reducing the waveguide width increases optical mode–sidewall interaction, amplifying scattering losses and reducing the intrinsic cavity Q. Conversely, wider (potentially multimode) waveguides achieve higher Qs but introduce unwanted spectral modes. To resolve this trade-off while pushing Q to higher numbers, we design a geometrically filtered single-mode (GFSM) microresonator by introducing tapered sections into a multimode microring (see Fig. 3(a)). Adiabatic tapers transition a small cavity section to single-mode, suppressing higher-order modes while maintaining low loss and high coupling efficiency [45]. Figure 3(a) shows a top-view scattering electron micrograph (SEM) of the GFSM design. This design retains a wide (width = 4.6 μm) multimode waveguide over most of the 2.7 mm peripheral length of the cavity (R = 340 μm) with four 70 μm-long tapers that reduce the width to 2.1 μm in localized regions for coupling to the bus waveguide and/or adjacent resonators with similar shapes. Two additional 130 μm-long single-mode sections facilitate critical coupling at larger gaps, improving fabrication tolerance and reducing perturbation while balancing compactness and FSR. Experimental results (Fig. 4) show FSR ~ 64.9 GHz and an average intrinsic $Q \sim 2.1 \times 10^7$ while demonstrating effective higher-order mode suppression and low-loss operation in the GFSM design.

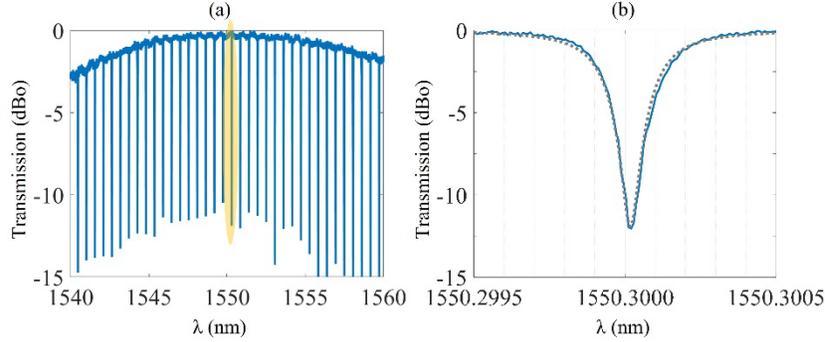

Fig. 2. (a) The spectral response of $R = 400$ μm single-mode microring resonator in the transverse-electric (TE) polarization (electric field parallel to the substrate plane). The waveguide width is 2 μm. The transmission envelope curve is due to bandwidth response of the input/output (I/O) gratings. (b) A specific resonance highlighted in (a) is zoomed in and further investigated to determine the cavity $Q$, for which the experimentally extracted curve (solid) is fitted to a Lorentzian function (dotted) confirming intrinsic $Q > 1.2 \times 10^7$. Note that the extinction exceeds 12 dB, confirming the design accuracy for critical coupling. Data for (b) is collected at smaller range of sweep for laser (500 pm total sweep between 1550.1 and 1551.6 nm) to provide a linear sweep and accurate $Q$ measurements [44].

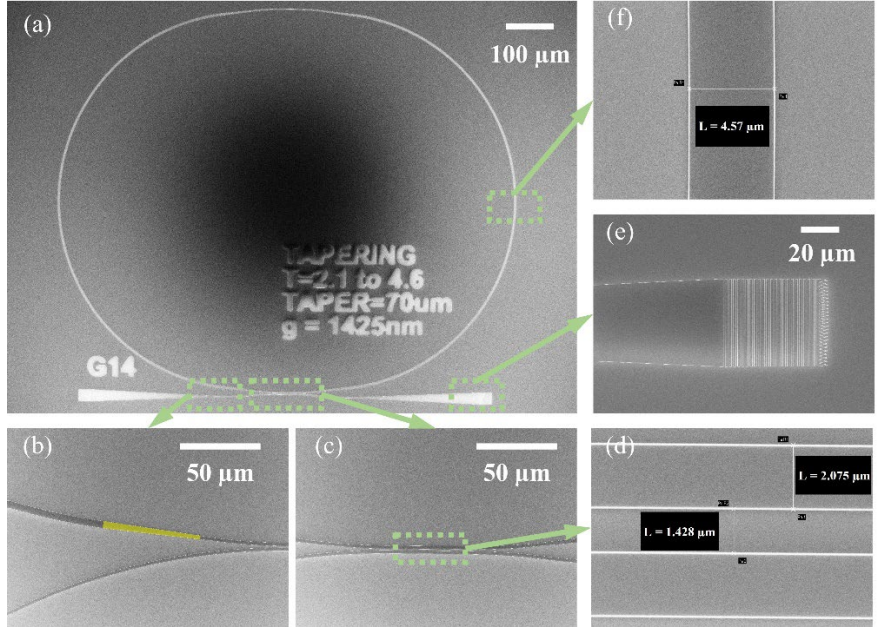

Fig. 3. (a) SEM of an all-pass single-pole filter (APF) fabricated on SiN. The center of the image is darker partly due to the inherent electron charge-up effect. (b) Enlarged image of the 70 µm-long taper section (yellow) in the cavity. (c)-(d) Enlarged images of the waveguide-to-cavity coupling region of the structure. The fabricated gap is measured at 1428 nm, which closely matches the design value of 1425 nm. Also, the waveguide width of 2075 nm follows the design value of 2100 nm within 2% margin. (e) Test structures such as APF are utilized with grating couplers to facilitate laser input/output coupling. (f) Enlarged image of the multimode waveguide section of the cavity.

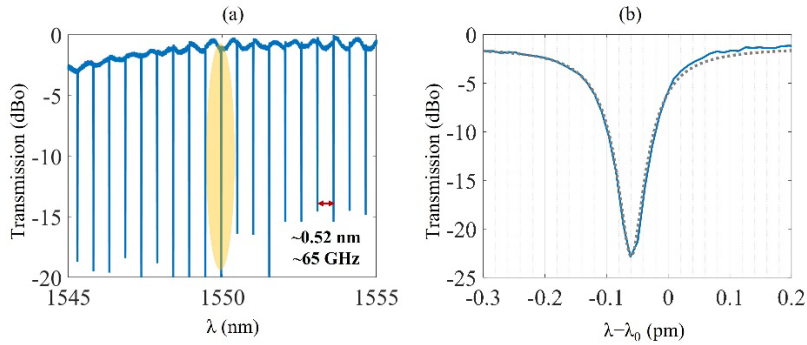

Fig. 4. (a) The spectral response for GFSM resonator in Fig. 3. The transmission envelope modulation is due to the presence of FP resonances within access waveguide as well as the input/output grating. (b) Zoomed version of the resonance highlighted in (a), for which the experimentally extracted curve (solid blue) is fitted to a Lorentzian function (dotted gray curve) confirming intrinsic $Q > 2.1 \times 10^7$.

## 5. Sustainable Filtering Operation and Tuning Mechanism of High-order Integrated Photonic Filters

Precise control of the spectral position of each of the five poles of the 5$^{th}$-order filter is critical for realizing a narrow-passband coupled-cavity filter. Each pole depends on the effective cavity length and inter-cavity coupling strengths [46,47]. While our design based on the Butterworth response formulation can nominally set these parameters, fabrication variations in length and width at the nanoscale perturb the effective index and resonance frequency and change the values of these poles. This is particularly problematic in high-Q resonators where resonance linewidths are much narrower than the filter bandwidth. Our calculations show that even a 42 nm (~0.0015%) length deviation in one cavity can add > 12 dB insertion loss (see Fig. S2). The tolerable deviation scales as ~$Q^{-1}$, requiring < 1 nm precision in the cavity length to keep resonance detuning within a single linewidth, thereby maintaining filter insertion loss below 3 dB. This is challenging to achieve, but it can be partially mitigated by identical cavity fabrication and rigorous process control. Temperature gradients across cavities must also remain < 100 mK, based on SiN and $SiO_2$ thermo-optic coefficients ($2.45\times10^{-5}$ $K^{-1}$ and $9.5\times10^{-6}$ $K^{-1}$, respectively [48]). To address these challenges, we integrate microheaters with each cavity to enable post-fabrication trimming and robust operation under ambient fluctuations. The heaters are fabricated via standard lift-off, with the heater areas and probe pads defined by EBL. Following fabrication, the devices are wire-bonded to allow electrical access for tuning. Figure S3 shows the final sample after wire bonding. Lower heater resistance is achieved using an interdigitated design, supporting CMOS-level control, which has been critical for compact and efficient controllers without requiring voltage amplification. For long-term stability, a derivative-free electrical-optical feedback algorithm is implemented, which is noise-tolerant and can lock cavities to a laser reference at low signal powers [49]. The controller uses a mathematical model of the device and monitors output power in a closed feedback loop to sustain multi-pole cavity alignment over time. Developing the control method requires defining an objective function, here chosen as maximizing the output power at the center wavelength. The objective function $T(\theta_1,\theta_2,\theta_3,\theta_4,\theta_5)$ represents the output intensity as a function of the phases $\theta_i$ applied to each microresonator to compensate for non-idealities and environmental variations. For a multipole filter, the function T is unimodal with few local maxima near the global maximum, ensuring convergence if the algorithm adequately samples the parameter space. A sampling-based search algorithm is appropriate due to its simplicity, noise resilience, and suitability for this optimization. Among options, the coordinate search method is practical for proof-of-concept demonstrations. It sequentially optimizes each phase $\theta_i$ by line search along one dimension at a time while holding others fixed, cycling through all dimensions iteratively. Although slower to lock, the method suffices as drift is minimal over short timescales once initial locking is achieved. Sustainable operation is thus feasible using coordinate ascent at fast timescales. Finally, this approach requires the microheaters to provide tuning over a range at least equal to the cavity's FSR to ensure convergence during control. Figure 5 illustrates a typical cavity-locking procedure, in which the initial state of cavities is far perturbed from the desired state of operation (Fig. 5(a)). With any random selection of initial state in the cavities with respect to each other, as well as the laser input, the method can lock into the desired state of operation (Figs. 5(b)-5(d)). While the initialization stage may be lengthy and take tens of iterations to accurately lock within ± 10 MHz (1% equivalent), it is important that the sustained operation occurs within much lower iterations, since the initial state and the desired state are close.

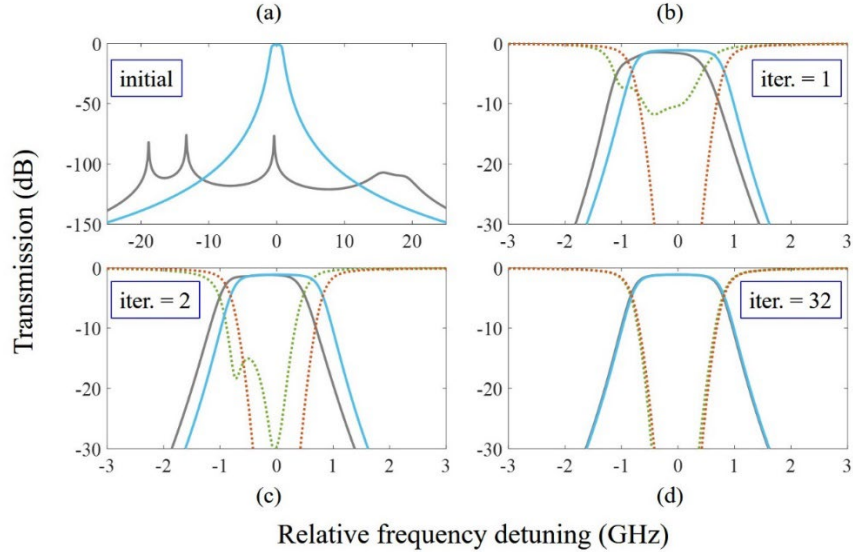

Figure 5. Numerical demonstration of the coordinate ascent method for laser-locking of the 5-pole coupled cavity filter, in which the initial state of the response (gray/dashed green) is adjusted through feedback to arrive at the desired spectral response (blue/dashed brown). In all figures, the gray (blue) curves correspond to the calculated (desired) response at the drop port, and the dashed green (brown) curves correspond to the calculated (desired) response at the through port. (a) The initial state of cavities is randomly generated resulting in a large difference between the resulting (gray) and the desired (blue) response. Here, the five cavities have an estimated detuning of -18, -13, -0.4, 15 and 19 GHz from the laser source. (b) Considerable improvement in drop port response through a single iteration is achieved, but the error in the through response is high. (c) Through port also achieves decent extinction in the second iteration. (d) The filter response is virtually identical to the designed spectral response after 32 iterations. The mismatch is less than 15 MHz, which is mainly due to assumption of 8-bit digitization over the control signal.

## 6.     Demonstration of Single- and Coupled-Cavity Integrated Photonic Filters

A critical metric for demonstrating coupled-cavity filters is the fidelity of design parameters—such as power-coupling ratios and pole center frequencies—relative to measured values. This compliance strongly impacts the spectral response but is challenging to verify in coupled cavities, since individual resonators cannot be directly accessed. To establish a baseline for higher-order filters, single-cavity band-pass and band-stop filters with varying bandwidths are first fabricated and characterized. Extensive experimental details of the single-pole and three-pole filters as the early stages of the final design are provided in the Supplementary materials (see Figs. S4 and S5 for a tunable single-pole filter with ~7.5 GHz 3 dB bandwidth at telecom wavelengths and Fig. S6 for a 3-pole device). Building on these demonstrations, we designed and fabricated a set of 5-pole filters with bandwidths ranging from 500 MHz to 5 GHz, incorporating low-resistance microheaters for active and independent control of each pole. Figure 6(a) shows the wafer during processing, while Fig. 6(b) illustrates the 5-pole integrated photonic filter based on the GFSM microresonator design after singulation and facet polishing. The polished chips are then wirebonded for thermal control of the poles. The resulting filters are characterized using edge-coupling of the fibers with the on-chip bus waveguide using a tunable laser. Figure 7(a) illustrates the measured spectral response of the 5-pole filter designed to achieve 1 GHz 3-dB passband. Building block GFSM microresonators each have ~2.7 mm roundtrip length that provides 65 GHz FSR, in close agreement with measured FSR = 64.5 GHz. As imperfections in fabrication perturb each cavity differently, post-fabrication active alignment is required using microheater elements. The fabricated 5-pole filter with a 1-GHz 3-dB passband consumes a total tuning power of 145 mW (<30 mW per cavity), indicating a low overall tuning overhead. The measured bandpass in Fig. 7(a) is 1.03 GHz, in close agreement with 1 GHz design (3% deviation). The ripple is measured at 0.4 dB thanks to the active control of cavities and fabrication compliance to the maximally flat design. The design offers an unprecedented >55 dB out-of-band rejection ratio due to employing the 5-pole design. In comparison, the single-pole demonstration can provide 16 dB (see Fig. S4). In Fig. 7(b), the spectral response of the 5-pole filter with 520 MHz target (or designed) passband is shown, in which the similar maximally flat response is established with < 0.1 dB ripple and > 55 dB out-of-band rejection ratio (measured up to -55 dB due to the limitation of the measurement system). The measured 3-dB passband is 481 MHz, for which there is nearly 8% deviation from the design. The increased deviation compared to the 1 GHz filter in Fig. 7(a) is deemed to be due to the higher sensitivity of the design to fabrication non-idealities as the bandwidth is reduced. Most importantly, the measured insertion loss (excluding input/output fiber-to-chip coupler loss) for both 1 GHz and 520 MHz filters are ~2 dB, which is, to the best of our knowledge, the lowest reported for an integrated photonic 5-pole filter with a spurious-free spectral response. The end-to-end coupling efficiency of the photonic integrated circuit with a similar waveguide and no optical filter is measured to be approximately 26% or -6 dB. We separately measured a fiber-to-waveguide coupling loss of ~2dB per fact by characterizing several integrated photonic waveguides with no filter. The response of the filter at one FSR away from the operation wavelength (at which the filter is optimally tuned) is demonstrated in Figure S10. Even with no additional tuning, the filter shape is approximately invariant across over 130 GHz. Thus, this filter has a wide range of tunability to meet center and intermediate frequency specifications.

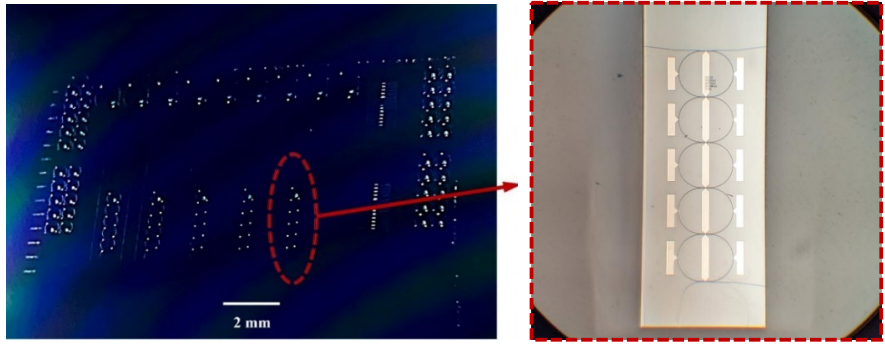

Fig. 6. (a) Optical micrograph of the wafer processed for integrated 5-pole filter with 1 GHz bandwidth. Inspection and quality control devices are co-fabricated to stablish performance metrics. (b) The micrograph of a 5-pole integrated filter utilizing the GFSM microresonator design. Input/output waveguides at chip facets are angled (7° to 8° with respect to substrate Si crystalline directions) to avoid back-reflection to the laser and remove stray-light coupling.

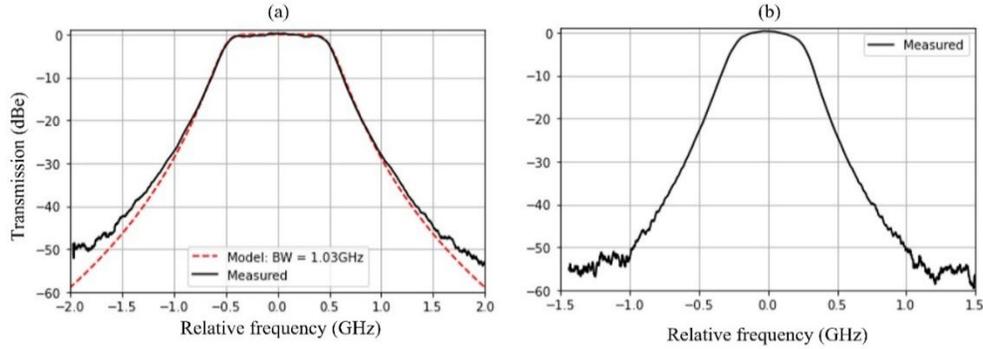

Fig. 7. (a) Measured response of the fabricated 5-pole integrated microring-based filter designed for the TE polarization. The curves represent the normalized experimental transmission response of the filter (signal-out, solid black) and the Butterworth filter model with 1.03 GHz 3-dB bandwidth (dashed red).. (b) Measured response of the fabricated 5-pole filter with 520 MHz target bandpass, showing 3-dB bandwidth of 481 MHz. Spurious-free bandwidth of 65 GHz is demonstrated thanks to the high FSR and the single-mode operation of the device.

## 7. Discussion

The demonstrated results confirm that the GFSM resonator design enables high-order coupled -cavity filters with a unique and record high combination of large FSR (~65-70 GHz), intrinsic Q (~2.1 × $10^7$) at this FSR range, and ultra-low insertion loss. The measured 2dB loss for both the 1 GHz and 520 MHz filters, represents the lowest reported to date for integrated 5-pole devices with spurious-free spectral response. The measured passbands closely match their design targets, with ripples below 0.4 dB, and the filter response remains invariant while tuning across more than 130 GHz without re-optimization. These results validate the design methodology and demonstrate the scalability of the SiN platform for realizing high-order, narrow-band integrated photonic filters. In addition, the modest tuning power (<30 mW per cavity) and stable thermal control establish the feasibility of sustained multi-pole operation, while the >55 dB out-of-band rejection ratio highlights the advantages of this approach for RF, microwave, and millimeter-wave applications.

Despite providing the superior filtering performance (thanks to the record-high Q at high FSR of the GFSM resonators), the demonstrated loss and Qs in this work have not fully met the potentials introduced in the design. For example, minimizing the length of the narrow-waveguide region in the GFSM resonator with any given peripheral length will further reduce the loss (as the waveguide loss reduces by increasing its width). Thus, a rigorous optimization of the GFSM structure for achieving highest Q with single-mode operation can further improve Q.

## 9. Conclusions

We have demonstrated a novel coupled-cavity filtering architecture enabled by the GFSM resonator architecture that enables single-mode operation while taking advantage of higher Qs in multi-mode architectures. The resulting combination of $Q \sim 2.1 \times 10^7$ and FSR ~70 GHz makes the demonstrated GFSM unique. The filters were designed by carefully engineering the coupling between adjacent resonators in the coupled-cavity structure to accurately model the poles of a 5$^{th}$-order Butterworth filter, with dynamic trimming of the resonances added by integrating thermal heaters with each resonator. The experimental demonstrations agree well with the theoretical designs and provide unprecedented combination of 2 dB insertion loss and >55 dB out-of-band rejection ratio at 1 GHz and 520 MHz bandwidths. The filter architecture is scalable, and the architecture can be extended to higher filter orders, making a reliable path for realization of manufacturable integrated photonic structures for tunable, low loss, narrow-band and frequency-stabilized filters that are of high demand for realization of transceivers for photonic front-ends, efficient signal processing in telecommunications, and sensing applications at the presence of strong interference and noise. In addition to electronic-photonic applications (e.g., for 5G and 6G telecommunication), broader utilization of such narrow-band and spurious-mode-free structures can be envisioned in nonlinear and quantum photonic applications (e.g., in optical pump filtering and separation of entangled states).

# Systematic Design and Demonstration of Multipole, Coupled-Cavity Integrated Photonic Bandpass Filters with High FSR, Single-Mode Microresonators in Low-Loss, Tunable Silicon Nitride Platform

Amir H Hosseinnia[1], Abdel Karim El Amili[2], Yu-Hung Lai[2], Danny Eliyahu[2], Mohammad Enjavi[1], Lute Maleki[2], Ali Adibi[1,*]

[1] School of Electrical and Computer Engineering, Georgia Institute of Technology, Atlanta, GA, 30332 USA
[2] OEwaves Inc., 465 North Halstead Street, Suite 140, Pasadena, CA 91107, USA

*ali.adibi@ece.gatech.edu

### S. 1. Trade-offs Among Performance Measures of the Coupled-cavity Butterworth Filter and the Effect of Resonator Q

Figure S1 of the manuscript clearly shows the importance of high-Q resonators in achieving a desired insertion loss at a given bandwidth. Considering other parameters of the device, a trade-off exists among the passband losses, the bandwidth, and the out-of-band rejection ratio. Any flexibility in one of the filter constraints can relax the minimum $Q$ required for the filter design, while a higher $Q$ can enable better performance at the expense of increased nanofabrication complexities to form better microcavities. We perform numerical simulations to study the minimum achievable bandwidth for a given intrinsic Q, examining device parameters and performance, while keeping the insertion loss at 50% (yielding ~50% optical power efficiency) for practical applications. Figure S1 shows the results of our simulations: bandwidths of 500 MHz, 650 MHz, 800 MHz, 1 GHz, and 1.5 GHz correspond to minimum required Qs of $3.65 \times 10^6$, $2.8 \times 10^6$, $2.2 \times 10^6$, $1.8 \times 10^6$, and $1.2 \times 10^6$, respectively. Notably, the minimum bandwidth varies approximately linearly with Q in this range. Figure S1 also clearly shows that higher Qs result in faster drop off of the filter response and higher out-of-band rejection ratios.

### S.2. Influence of Non-idealities on the Performance of an Optimized 1 GHz 5-pole Butterworth Bandpass Filter

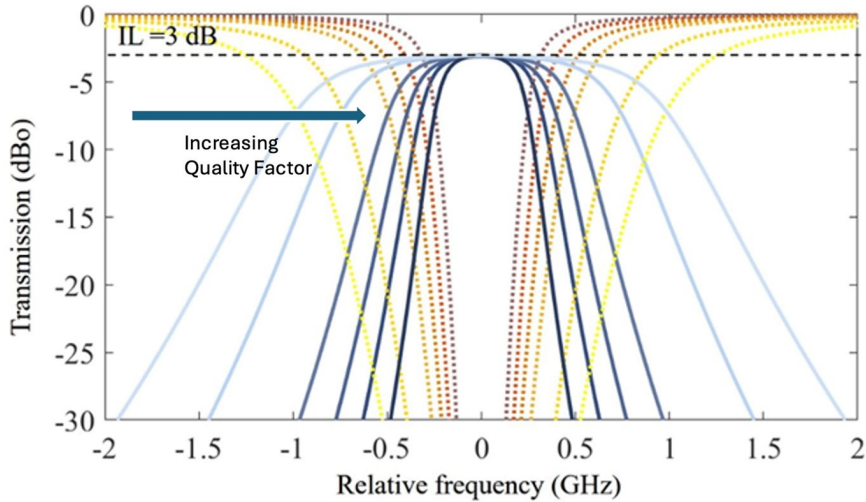

Fig. S1. Optical power transmission spectra for an N=5 coupled-cavity Butterworth filter (solid lines: drop port, dashed lines: through port) derived for minimum $Q$ required for various bandwidths while fixing the insertion loss at 50% (IL = 3dB). Here, the bandwidth is tightened from 1500 MHz to 500 MHz (light blue to dark blue in the drop port, and equivalently yellow to brown in the through port), which necessitates the $Q$ to increase from $1.2 \times 10^6$ to $3.65 \times 10^6$. For bandwidths of 1 GHz, 800 MHz, and 650 MHz, the $Q$ values are derived as $1.8 \times 10^6$, $2.2 \times 10^6$, and $2.8 \times 10^6$, respectively.

Figure S2 illustrates the sensitivity of the optimized 1 GHz, 5-pole bandpass photonic filter to cavity length variations. A deviation as small as 42 nm—only ~0.0016% of the original 2.7 mm cavity length—can significantly distort the spectral response and introduce more than 12 dB additional insertion loss (light blue curve). To model fabrication and thermal non-idealities, one of the five cavities was perturbed by 42 nm, 28 nm, and 14 nm, with the corresponding filter responses shown in progressively lighter to darker blue. Even these small deviations shift the spectral shape, degrade the flat passband, and considerably increases the insertion loss. For comparison, the unperturbed filter is also shown, with the drop-port response in dark blue and the through-port response in dashed yellow. These results underscore the need for precise fabrication control and robust thermal stabilization to realize high-order integrated photonic filters.

### S.3. Thermal Control of Filter Poles via Wirebonding of Polished Chips
The polished chips were mechanically mounted on a test fixture and wirebonded to enable

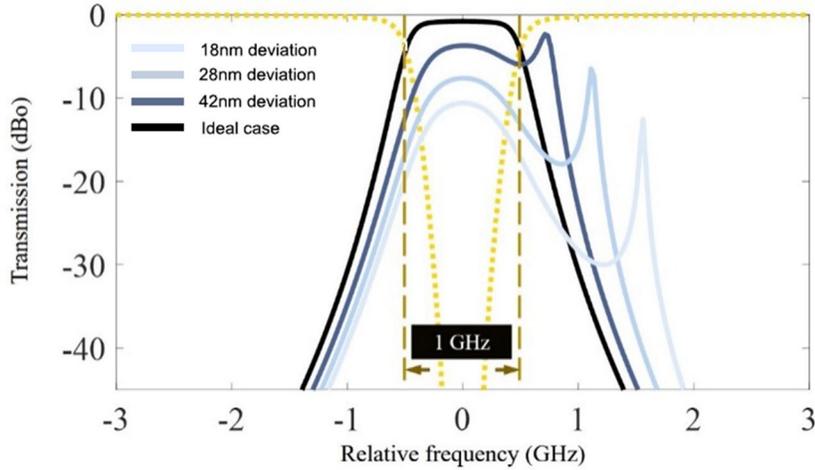

Fig S2. Variation of the filtering performance of an optimized 1 GHz 5-pole bandpass integrated photonic filter due to fabrication non-idealities or environmental variations. One of the 5 cavities is perturbed to have 42, 28, and 14 nm deviation in length (lighter to darker blue, respectively) from the optimally designed 3 mm value. The perturbation can occur due to fabrication imperfections or thermal variations over the chip. For comparison, the unperturbed curve is also depicted (dark blue: drop port, dashed

thermal control of the poles, as shown in Fig. S3.

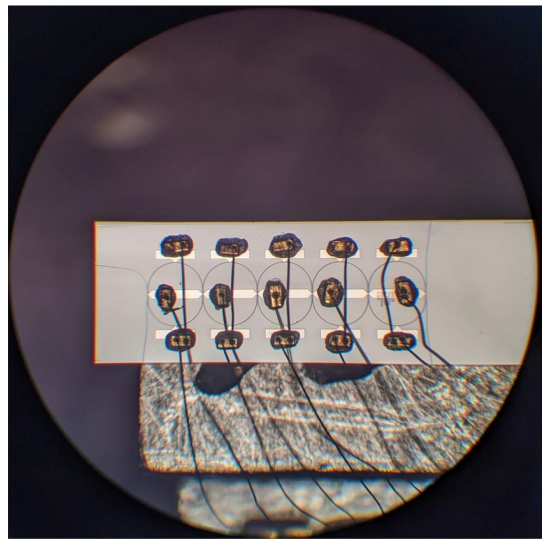

Fig S3. Optical microscope image of the fabricated 5-pole integrated photonic filter after singulation, facet polishing, and wirebonding.

## S.4. Design, Fabrication, and Testing of Single-Pole Filters

Figure S4 demonstrates the design and performance of a single-pole integrated photonic filter based on the GFSM microresonator architecture. The SEM image in Fig. S4(a) shows the fabricated device, which includes four access ports used for signal input/output and monitoring. To ensure efficient fiber-to-chip coupling, the waveguide tip is tapered to expand the guided optical mode to ~2 µm, closely matching the fiber mode field diameter (MFD). This tapering, combined with a 2 µm bus waveguide width, guarantees single-mode operation across the wavelength range of interest. The cavity–waveguide gap is set at 1450 nm to provide the desired coupling strength for the filter design. The measured power transmission spectra, plotted in Fig. S4(b), verify the intended single-pole filter response in TE polarization. The drop-port transmission (solid blue) and through-port monitor (dashed gray) confirm the filter's designed 7.5 GHz bandwidth and FSR = 70 GHz (experimentally measured FSR ~ 68 GHz). To further assess the filter fidelity, Fig. S4(c) enlarges one resonance, revealing that the experimental results (solid curves) are in near-perfect agreement with numerical simulations (dashed curves). This strong correspondence validates both the designed cavity–waveguide gap and the chosen power coupling ratio, demonstrating that the GFSM-based single-pole filter provides precise spectral shaping with minimal deviation from theoretical predictions.

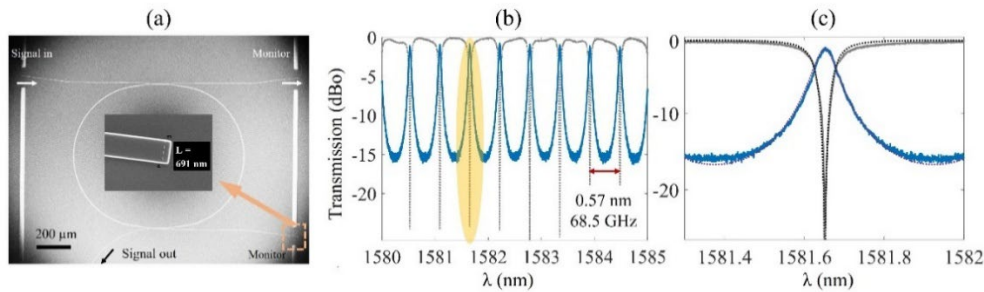

Fig. S4. (a) SEM image of the single-pole GFSM microresonator filter with 7.5 GHz bandwidth. Inset: tapered waveguide tip matched to 2 µm MFD; cavity–waveguide gap is 1450 nm. (b) Measured transmission spectra of drop (solid blue) and through (dashed gray) ports under TE polarization with FSR~ 68 GHz. (c) Enlarged resonance showing close agreement between experimental (solid) and simulation (dashed) results.

Figure S5 illustrates the experimental setup used for coupling to the single-pole integrated photonic filter. As shown in Fig. S5(a), the input optical signal is launched into the chip using an anti-reflection (AR)-coated lensed fiber with a 2.5 µm mode diameter, providing efficient coupling to the tapered waveguide tips of the device. On the output side, large-core multimode fibers are employed to collect the transmitted signals, thereby minimizing collection loss and ensuring robust signal acquisition. For preliminary alignment, a red laser source was coupled into the input port, as presented in Fig. S5(b). Under a microscope, scattered red light becomes visible at the output and monitor ports, serving as a clear indicator of optimal fiber-to-waveguide alignment. This method allows for efficient alignment of the input and output coupling paths before switching to the intended operating wavelength for device characterization.

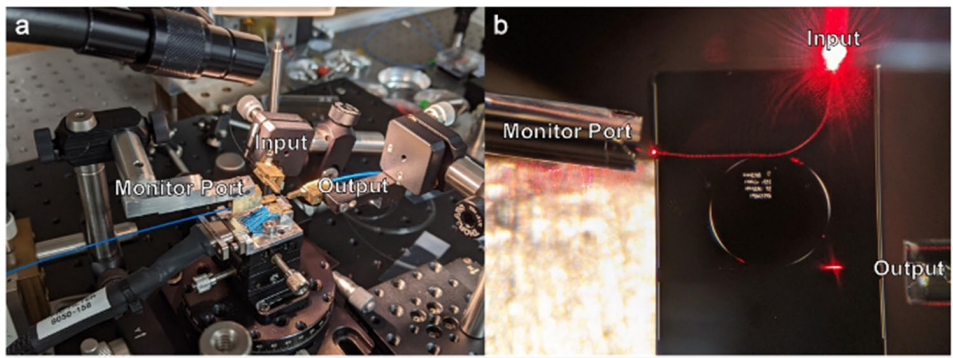

Fig. S5. Coupling setup for the single-pole integrated photonic filter. (a) Overview of the test station with the input to the chip provided via an AR-coated lensed fiber and the output collected by a large-core multimode fiber. (b) Preliminary alignment using red light, where scattered light under the microscope confirms optimized coupling at the monitor and output ports.

## S.6. Design, Fabrication, and Testing of 3-Pole Filters

A set of 3-pole integrated photonic filters are designed and fabricated for detailed characterization of the multipole filter designs, response tuning by co-integration of microheaters with cavities, and our developed active control method on the packaged chip. Such step-by-step approach enables point-by-point monitoring and inspection of control mechanism

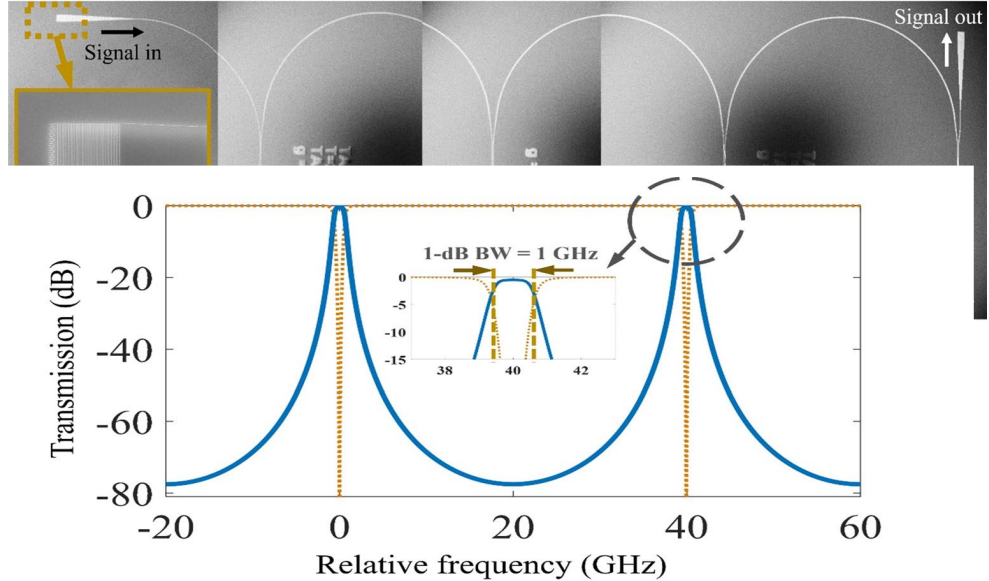

Fig. S6. Optical power transmission spectra of a third-order Butterworth filter numerically computed for minimum device insertion loss while fixing the 1-dB pass band at 1 GHz. Blue (solid) and brown (dashed) curves represent drop and through port, respectively. The inset also shows the in-band response at 1 GHz. Out-of-band rejection ratio above 70 dB is achieved using 3 poles, in part due to the high FSR employed in in reduced dimensions and subsequently guarantee success of higher-order designs with higher degrees of freedom in control. Figure S6 shows the computed spectral response of the 3-pole filter designed with FSR = 39.94 GHz at the telecommunication wavelengths. Following the design principles of a maximally flat Butterworth filter to realize a 3-pole integrated device, the couplings are calculated as $\kappa_1 = \kappa_3 = 0.34$ and $\kappa_2 = 0.041$. Figure S7 shows the top-view SEM of the fabricated device. The design of this 3-pole device utilizes tapered single-mode microresonators. Monitor ports are included in the design, which can be used to independently align input and output fibers, as well as providing complementing data for characterization of filter devices.Figure S8 shows the characterization result for the coupled microcavity at the output signal port. The main observation is the spectral splitting among degenerate modes of 3-pole microresonator structure. While the cavities are identical, it is clear from the characterized device that the realized frequency of each cavity is different in the resultant multipole device. The spectral splitting is due to the large difference in the couplings at each cavity, which leads to different effective group indices in the coupling region of cavities and corresponding

frequency offsets. This results in the double peak in Fig. S8(b). As a result, the insertion loss is high due to the spectral misalignment, and the measured spectral out-of-band rejection is therefore limited to the noise floor of the measurement setup. Both issues will be properly addressed by integration of active controller mechanisms (e.g., by using the thermal tuning) and tested in the upcoming set of active on-chip multipole devices.

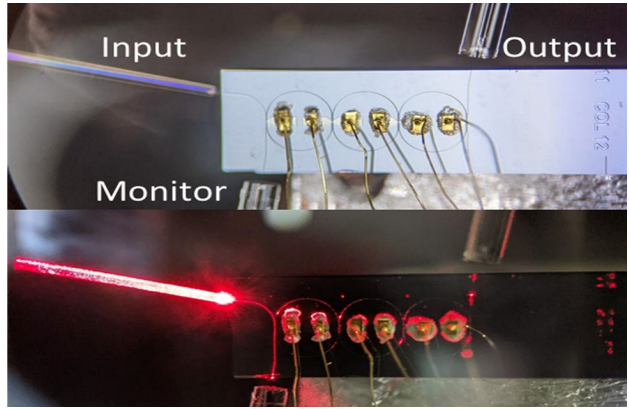

Fig. S9. Test setup for the 3-pole filter. Top: input coupling via a lensed fiber with lightpipes collecting the monitor and output signals. Bottom: red laser used for preliminary alignment of the input waveguide.

Figure S9 shows the experimental arrangement for testing the 3-pole filter device. As illustrated in the top panel, the optical input is coupled into the photonic chip using a lensed fiber, while the outputs at both the monitor and signal ports are collected using lightpipes to ensure efficient extraction of the transmitted light. For accurate alignment, a red laser was first coupled into the device, as shown in the bottom panel. The visible scattered red light under a microscope provides direct feedback on fiber–waveguide alignment, enabling precise positioning of the lensed fiber before switching to the operating wavelength for filter measurements.

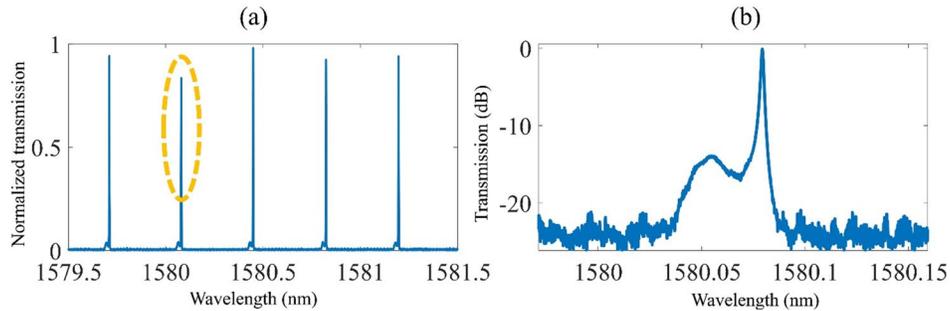

Fig. S8. (a) The transmission spectrum of the 3-pole filter, and (b) the zoomed response of filter at the specific wavelength highlighted in (a). The measured FSR = 40.27 GHz is well preserved within 1% of the design value (FSR = 39.94).

### S.7. Wideband Tunability and Invariant Filter Response

The filter response is optimized for an input wavelength of 1551.2 nm. To evaluate its robustness, the input wavelength is swept around 1550.6 nm and 1551.6 nm (± one FSR away from the designed center wavelength) without applying any additional filter tuning. The corresponding spectra in Fig. S10 confirms that the filter maintains its shape over this detuning range, demonstrating stable performance across adjacent channels.

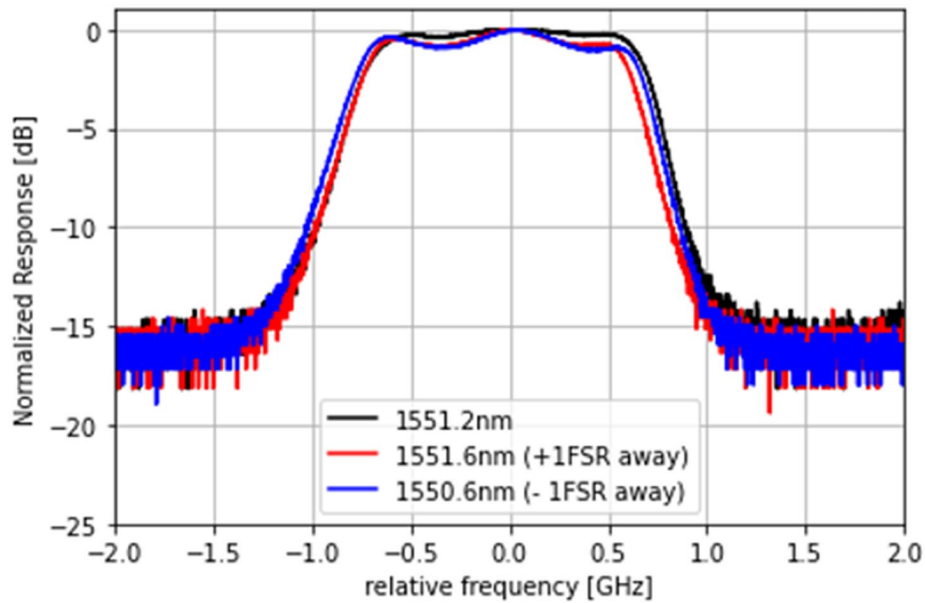

Fig. S10. Filter response optimized for input wavelength of 1551.2 nm, and corresponding spectra when the input wavelength is shifted by one FSR on both sides without tuning the filter.